\documentclass[10pt,conference,final]{IEEEtran}
\PassOptionsToPackage{bookmarks=false}{hyperref}

\IEEEoverridecommandlockouts
\IEEEaftertitletext{\vspace{-1\baselineskip}}

\usepackage[shrink=90]{microtype}
\usepackage{pifont}
\usepackage{tikz} 
\usepackage{booktabs}
\usepackage{booktabs}
\usepackage{dcolumn}
\usepackage{array}
\usepackage{listings}
\usepackage{courier}
\usepackage{flushend}
\usepackage{enumitem}
\usepackage[skip=6pt]{caption}
\usepackage[caption=false,font=footnotesize]{subfig}
\usepackage{amssymb} 
\usepackage{mathtools} 
\usepackage[bookmarks=false]{hyperref}
\usepackage{float}
\usepackage[utf8]{inputenc}
\usepackage{mdframed} 
\usepackage{listing} 
\usepackage[hang,flushmargin]{footmisc} 
\usepackage[T1]{fontenc} 
\usepackage{textcomp} 
\usepackage{colortbl} 

\def\BibTeX{{\rm B\kern-.05em{\sc i\kern-.025em b}\kern-.08em
    T\kern-.1667em\lower.7ex\hbox{E}\kern-.125emX}}
    
\usepackage[printwatermark]{xwatermark}
\usepackage{xcolor}
\usepackage{tikz}
\newsavebox\mybox
\savebox\mybox{\tikz[color=black,opacity=1]\node[text width=15cm,align=center]{Accepted at IEEE BigData 2020. For personal use only. Do not distribute.\\[.5em] T. Taylor, F. Araujo and X. Shu, ``Towards an Open Format for Scalable System Telemetry,'' 2020 IEEE International Conference on Big Data (Big Data), Atlanta, GA, USA, 2020, pp. 1031-1040.};}
\newwatermark*[
 page=1,
 angle=0,
 scale=.75,
 xpos=0,
 ypos=130
]{\usebox\mybox}

\newif\ifinc
\incfalse  

\newcommand{\pname}{{{SysFlow}}}

\newcommand{\inlinedsection}[2][6pt]{\vspace{#1}\noindent\textbf{#2}.}

\newcommand{\inlinedsectionit}[2][3pt]{\vspace{#1}\noindent\textit{#2}.} 

\newcolumntype{P}[1]{>{\centering\arraybackslash}p{#1}}

\newcommand\Cplusplus{C\raise1pt\hbox{\small++}}

\newcommand{\code}[1]{{\small \ttfamily #1}}
\newcommand{\tcode}[1]{{\scriptsize \ttfamily #1}}

\newcommand*\conj[1]{\overline{\mbox{$#1$\vphantom{\textit{p}}}}}

\newcommand{\tbvar}[1]{\raise.5pt\hbox{\scriptsize{(#1)}}}
\newcommand\innersep{\hskip3pt}
\newcommand\outersep{\hskip18pt}
\newcommand\outersepsmall{\hskip6pt}

\mdfsetup{skipabove=0pt,skipbelow=0pt}

\newcommand{\bchoice}{\mid}
\newcommand{\choice}{\bchoice}
\newcommand{\defn}{:=}

\newcommand{\grsymbol}[1]{{\textit{#1}}}
\newcommand{\production}{::=}

\newcommand{\langkw}[1]{\text{{\ttfamily #1}}}

\newcommand{\tmatch}{\langkw{match}}
\newcommand{\tshow}{\langkw{show}}
\newcommand{\ttag}{\langkw{tag}}
\newcommand{\twith}{\langkw{with}}
\newcommand{\texists}{\langkw{exists}}
\newcommand{\tin}{\langkw{in}}
\newcommand{\tpmatch}{\langkw{pmatch}}
\newcommand{\tcontains}{\langkw{contains}}
\newcommand{\tstartswith}{\langkw{startswith}}

\newcommand{\opnegt}{\langkw{not}\hspace{0.1em}}

\newcommand{\bor}{\langkw{or}}
\newcommand{\band}{\langkw{and}}

\newcommand{\policy}{\grsymbol{P}}
\newcommand{\flist}{\grsymbol{list}}
\newcommand{\fmacro}{\grsymbol{macro}}
\newcommand{\frule}{\grsymbol{rule}}
\newcommand{\cond}{\textit{c}}
\newcommand{\expr}{\textit{e}}
\newcommand{\binop}{\Diamond_b}
\newcommand{\unop}{\Diamond_u}
\newcommand{\term}{\textit{t}}
\newcommand{\var}{\grsymbol{v}}
\newcommand{\val}{\grsymbol{u}}
\newcommand{\op}{\textit{a}}
\newcommand{\sysflow}{\textit{sf}}

\def\Plus{\texttt{+}}
\def\Star{\texttt{*}}
\newcommand*\mult[1]{{{#1}^{\Plus}}}
\newcommand*\multopt[1]{{{#1}^{\Star}}}

\begin{document}

\title{\huge Towards an Open Format for Scalable System Telemetry}

\author{\IEEEauthorblockN{Teryl Taylor\textsuperscript{$\dagger$}\thanks{$\dagger$Both authors contributed equally to this research.}}
\IEEEauthorblockA{\textit{IBM Research} \\
terylt@ibm.com}
\and
\IEEEauthorblockN{Frederico Araujo\textsuperscript{$\dagger$}}
\IEEEauthorblockA{\textit{IBM Research} \\
frederico.araujo@ibm.com}
\and
\IEEEauthorblockN{Xiaokui Shu}
\IEEEauthorblockA{\textit{IBM Research} \\
xiaokui.shu@ibm.com}
}

\IEEEpubid{\copyright~2020 IEEE}

\advance\baselineskip-.56pt plus.1pt minus.5pt
\hyphenpenalty=5
\exhyphenpenalty=0

\maketitle


\begin{abstract}
A data representation for system behavior telemetry for scalable big data security analytics is presented, affording telemetry consumers comprehensive visibility into workloads at reduced storage and processing overheads. The new abstraction, SysFlow, is a compact open data format that lifts the representation of system activities into a flow-centric, object-relational mapping that records how applications interact with their environment, relating processes to file accesses, network activities, and runtime information. 
The telemetry format supports single-event and volumetric flow representations of process control flows, file interactions, and network communications. Evaluation on enterprise-grade benchmarks shows that SysFlow facilitates deeper introspection into attack kill chains while yielding traces orders of magnitude smaller than current state-of-the-art system telemetry approaches---drastically reducing storage requirements and enabling feature-filled system analytics, process-level provenance tracking, and long-term data archival for cyber threat discovery and forensic analysis on historical data.
\end{abstract}


\section{Introduction}
\label{s:intro}
Conventional perimeter monitoring offers limited coverage and visibility into workloads and attack kill chains, posing severe asymmetry challenges for defenders and impeding timely security analysis and response to resist known and emerging cyber threats. To cope with this visibility gap, comprehensive telemetry probes have recently been proposed to monitor system events and detect workload misbehaviors~\cite{LinuxAudit:2006, Sysdig:2019, Kcal:2018, protracer:2016}. However, current solutions expose system event information as low-level operating system abstractions (e.g., system calls), generating massive amounts of data and
making it impractical and prohibitively expensive to store, process, and analyze such telemetry data,
thus limiting analytics to rule-based approaches.

Existing system telemetry representations also suffer from unnecessary generalizations, trading interpretability and performance for the support of optional and custom attributes. This makes data schemas overly complex to ingest and process in realtime. Moreover, data formats lack support for concurrently tracking event provenance while being streamed live---critical for security investigations and root cause analysis, and most endpoint telemetry systems use proprietary data models, which impedes the implementation of standardized telemetry pipelines, forcing administrators to source monitoring data from multiple agents to satisfy regulatory requirements. 

To overcome these disadvantages, we propose \pname{} as a new data representation for system behavior introspection for scalable analytics. \pname{} is a compact \emph{open}\footnote{\url{https://github.com/sysflow-telemetry}.} 
data format that lifts the representation of system activities into a flow-centric, object-relational mapping that records how applications interact with their environment.
It connects process behaviors to network and file access events, providing a richer context for analysis. This additional context facilitates deeper introspection into host and container workloads, 
and enables a stream of cloud workload protection use cases, including system runtime integrity protection, process-level provenance tracking, and long-term data archival for threat hunting and forensics.

\pname{} captures information about host, container, process, file, and network relationships and activities, supporting \emph{single-event} and \emph{volumetric flow} representations of process control flows, file interactions, and network communications. 
\pname{} reduces data collection rates by orders of magnitude relative to existing system telemetry sources, and lifts events into behaviors that enable forensic applications and more comprehensive analytic approaches. Furthermore, the open serialization format and libraries enable integration with open-source data science frameworks and custom analytics. This simplifies big data pipelines for systems akin to the way NetFlow~\cite{Cisco:2012} did for network analytics. 

We also present a data processing pipeline built atop \pname{}. The pipeline provides a set of reusable components and APIs that enable easy deployment of telemetry probes for host and container workload monitoring, as well as the export of \pname{} records to S3-compliant object stores feeding into distributed security analytics jobs based on Apache Spark.
Specifically, the analytics framework provides an extensible policy engine that ingests customizable security policies described in a declarative input language, providing facilities for defining higher-order logic expressions that are checked against \pname{} records. This allows practitioners to easily define security and compliance policies that can be deployed on a scalable, out-of-the-box analysis toolchain while supporting extensible programmatic APIs for the implementation of custom analytics algorithms---enabling efforts to be redirected towards developing and sharing analytics, rather than building support infrastructure for telemetry.

\pname{}'s collection probe has been optimized to incur minimal performance overheads and does not require program instrumentation or system call interposition for data collection, therefore having a negligible impact on monitored workloads. The implementation has been validated under multiple stress test profiles. We describe its components in detail and demonstrate that the ability to continuously collect and store comprehensive system event information is critical for the identification of advanced and persistent threats, security vulnerabilities, and performing threat hunting tasks.

\IEEEpubidadjcol
Our contributions can be summarized as follows:
\begin{itemize}[leftmargin=15pt,itemsep=0pt]
    \item A compact data format and new flow analysis that records the interactions of processes and containers with their environment, providing an object-relational structure that enables many security and systems monitoring applications.
    \item A data processing language and pipeline for consuming and analyzing \pname{} records at scale.
    \item A capability analysis of \pname{} on its ability to express attack tactics, techniques, and procedures described in the MITRE ATT\&CK framework~\cite{Mitre:2019}.
    \item A comprehensive evaluation of state-of-the-art system telemetry solutions based on popular, enterprise-grade container applications and standard industry benchmarks.
\end{itemize}
\IEEEpubidadjcol
\section{Overview}
\label{s:overview}

\subsection{Computer Telemetry}
\label{ss:background}
Computer telemetry can be broadly categorized into network and system telemetry.
Network telemetry involves installing a passive tap at the perimeter of the network or on an endpoint to collect packet data. System telemetry includes application logging, program tracing, system call collection, and runtime metric gathering at endpoints. 

\inlinedsectionit{Network Telemetry} The excessive cost of full packet capturing has led researchers to develop techniques to reduce its storage burden. Maier et. al~\cite{Maier:2008} found that several days of traffic could be stored by collecting raw PCAP (Packet Capture) data up to a configurable byte cutoff per connection while retaining a complete record for the majority of network connections. This \textit{time machine} was combined with the Bro (Zeek) intrusion detection system~\cite{Paxson:1999} to enable real-time streaming analytics for security and retroactive data enrichment.  

While storing packet headers has also been used to reduce storage requirements, the most popular network telemetry system is NetFlow~\cite{Cisco:2012}, which aggregates packets from network sessions into single records that contain the 5-tuple identifier of the sessions combined with volumetric  information. NetFlow provides orders of magnitude compression over packet collection and has been a staple for network-based analytics for the past two decades. Extensions have been proposed to improve the format~\cite{Li:2016,Estan:2004}, augmenting it with more attributes~\cite{Taylor:2012}, and building more scalable probing mechanisms~\cite{Deri:2003}. 

However, despite its popularity, network telemetry only provides \textit{partial} visibility into system workloads, requiring a host-level monitoring source for capturing system behaviors.

\inlinedsectionit{System Telemetry}
While research on network telemetry has focused on data representations for scalable storage and processing, system telemetry is still dominated by approaches that operate on raw system call data~\cite{LinuxAudit:2006, Sysdig:2019, Kcal:2018, protracer:2016, Spade:2012, Rain:2017, Oliveira:2017, Lee:2013, Madhumathi:2018}. This causes many issues in terms of data collection size, efficiency, and performance. To alleviate these challenges, KCAL~\cite{Kcal:2018} improves data collection speeds through a kernel cache and aggressive system call filtering, while LogGC~\cite{Lee:2013} leverages a post-processing algorithm for pruning unreachable objects. SPADE~\cite{Spade:2012} defines a graph model for data provenance collection and storage, and ProTracer~\cite{protracer:2016} combines both logging and unit-level tainting to achieve cost-effective provenance tracing. Fmeter~\cite{Marian:2012} reduces data pressure by creating feature vectors of system call counters to perform indexable searches. System call analysis is also used in debugging for deterministic process replay~\cite{Chen:2015,Rain:2017}, 
which typically requires the injection of special libraries to record time and event features~\cite{Geels:2006}.

Among the most popular system call monitors is Linux Audit~\cite{LinuxAudit:2006}, which has been the de facto standard for GNU/Linux tracing since Kernel version 2.6, being used by many Linux distributions, and tools such as osquery~\cite{osquery}. Our evaluation (\S\ref{s:evaluation}) shows that Audit is not ideal for comprehensively collecting system call data at cloud scale---its kernel probe looses events under load, telemetry output is too large, and native container support is not yet upstream. Sysdig~\cite{Sysdig:2019} scales better than Audit for collecting system call information; however, it yields significant data footprints, making it intractable to store the telemetry stream, and therefore reducing data processing capabilities to low-overhead rule-based techniques.
The telemetry exported by these monitors also make it extremely difficult to efficiently build event provenance graphs or perform cyber threat hunting. To overcome these issues, commercial endpoint monitoring solutions
adopt their own proprietary formats. However, this requires the installation of multiple monitors to support various security products, wasting important cloud computing resources, and making integration across security products difficult. 

\inlinedsectionit{Limitations in Data Representation}
While many existing endpoint telemetry solutions provide low-level system information that can be harnessed for attack reasoning, provenance tracking, and performance analysis, they often suffer from inefficiencies in tracing~\cite{MPI:2017, partition:2013} and excessive data redundancy~\cite{Kcal:2018}, exacerbating the challenges of collecting, storing, and processing such telemetry data.
Several techniques have emerged in an attempt to mitigate data requirements, including folding sub-groups of system events according to their semantic meaning~\cite{HFD:2016, NodeMerge:2018} and pre-processing events at the probe level to reduce data redundancy~\cite{Kcal:2018}. However, none offer a comprehensive telemetry specification that satisfies the performance and data representation requirements of host-level telemetry.

Most recently, a new ontology for system telemetry has been developed in the context of the DARPA Transparent Computing (TC) program~\cite{tc:2014}, which investigated how security analysts would benefit given access to a \emph{complete} and \emph{high-fidelity} recording of computations across every endpoint in a compromised computer network. The program brought together teams from academia and industry to work on the generation and analysis of real-time monitoring of computations on an array of system environments and architectures. TC defined a Common Data Model (CDM) for communicating system behaviors and for representing data provenance and information flow semantics based on system calls. 

However, despite its many innovations on system monitoring~\cite{Rain:2017, Kcal:2018, CADETS:2017}, cyber reasoning methodologies~\cite{TIC:2018}, and automated analytics~\cite{Holmes:2019,Sleuth:2017}, TC engagements were conducted in small networks. As originally envisioned, TC does not scale to large enterprise environments due to the sheer volume of fine-grained telemetry data it generates, and the difficulties to transport, ingest, process, and store data in CDM, which is not designed for compactness. In addition, CDM creates excessive processing complexity through the support of hundreds of optional attributes, allowing it to be interpreted differently by various users. CDM advances system telemetry but has limited support for efficient stream processing for real-time security and compliance analysis in large-scale deployments.

\subsection{System Telemetry Requirements}
\label{ss:telereq}

The lessons we learned through the development of a cyber reasoning platform~\cite{TIC:2018} for DARPA TC informed many practical requirements for effective system telemetry. In particular, our experiences with system event analysis and the CDM format revealed that 
(1)~system telemetry is often collected at system call granularity---making it impractical to scalably process and store historical telemetry data,
(2)~employed schemas offer limited support for data provenance; collection stacks do not support provenance tracking and data streaming simultaneously, and (3)~existing data representations trade interpretability for flexibility through optional and custom attributes, hindering processing performance. In addition, telemetry systems often use proprietary data models, forcing administrators to source telemetry data from multiple agents, wasting resources, and making integration more challenging; this issue is exacerbated by the lack of open interfaces for consuming the collected data.

To cope with these shortcomings, an efficient system telemetry approach must offer a combination of lightweight, non-intrusive event collection with a data representation that is open, non-redundant, compact, and capable of expressing essential system properties and behaviors. Such representation must also abstract away noisy low-level system events while preserving the semantics of system executions, to enable a telemetry pipeline that is conducive of dense, long-term archival of system traces, and efficient reasoning over historical system behaviors. Furthermore, the format must inherently support linkages of system behaviors with processes, users and containers to enable data provenance and process control flow graphs on streaming data.    

\subsection{Towards System Flows}
\label{ss:motivation}

To achieve these goals, we envision a system telemetry stack that is capable of processing and summarizing system events into process behaviors while preserving system execution semantics, analogous to how NetFlow aggregates raw packet information. Towards this end, we introduce \pname{} as a new flow-based data specification for system monitoring. NetFlow played a pivotal role in scaling network-based analytics by condensing packet information into a much smaller flow data; furthermore, Netflow was widely adopted by the community because it is an open source format, and simple to use.   \pname{} adapts and extends such flow abstraction to model systems entities and interactions, collecting metadata from system calls and grouping sequence of events sharing the same properties.


\begin{figure}[t]
     \centering
     \includegraphics[width=1\hsize]{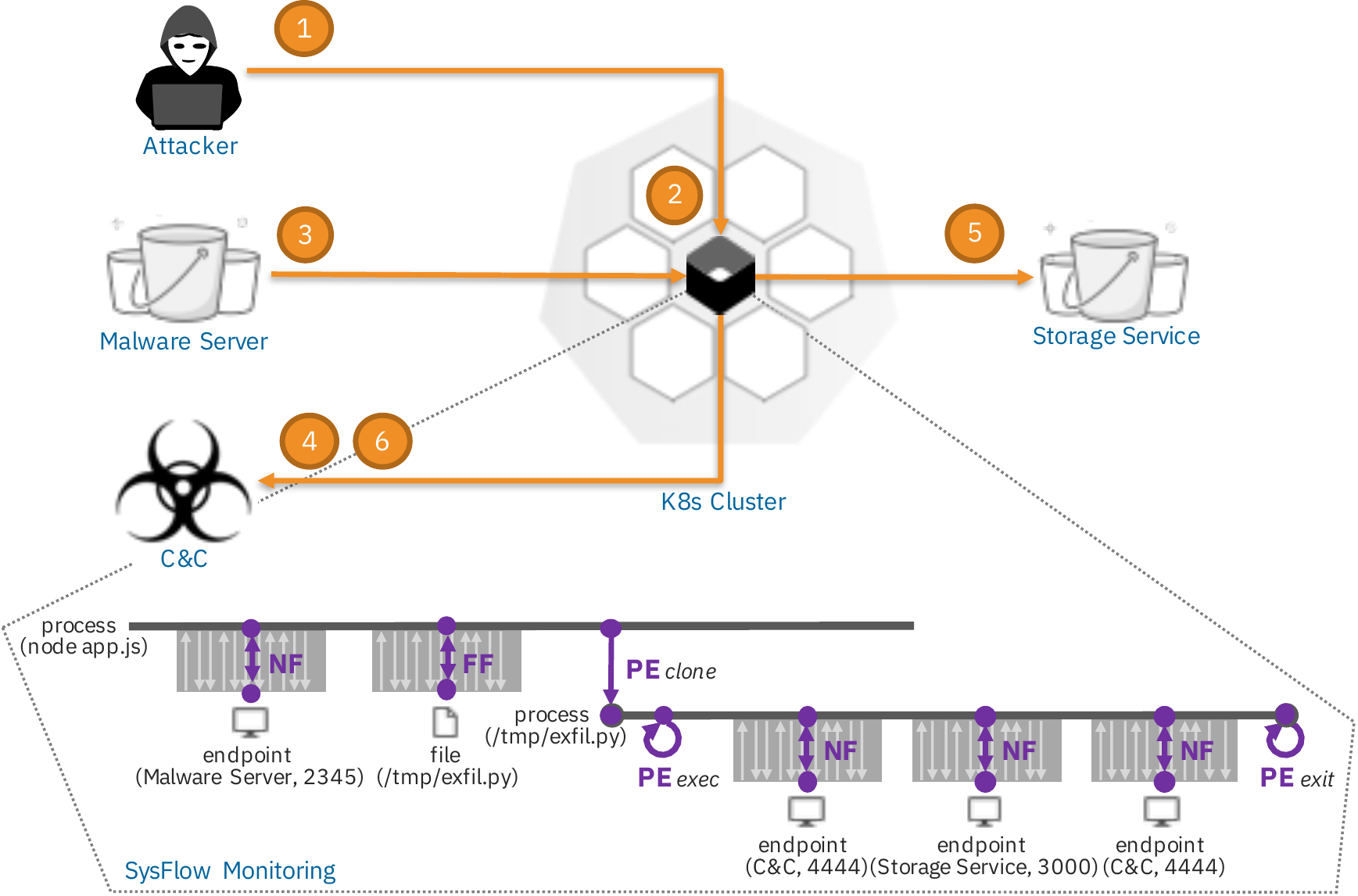}
     \caption{Using \pname{} to uncover a targeted attack.}
     \label{fig:overview}
 \end{figure}

Figure~\ref{fig:overview} shows how \pname{} can be used to uncover a targeted attack in which a cyber criminal exfiltrates data from a cloud-hosted service. During reconnaissance (step 1), the attacker detects a vulnerable node.js server that is susceptible to a remote code execution attack exploiting a vulnerability in a node.js module (e.g., CVE-2017-5941~\cite{CVE:2017:5941}). The attacker exploits the system using a malicious payload (step 2), which hijacks the node.js server and downloads a python script from a remote malware server (step 3). The script contacts its command-and-control server (step 4), and then starts scanning the system for sensitive keys, eventually gaining access to a sensitive customer database (step 5). The attack completes when data is exfiltrated off site (step 6).

While conventional monitoring tools would only capture streams of disconnected events, \pname{} can connect the entities and effects of each attack step on the system. For example, the highlighted \pname{} trace records and reveals precisely the steps in the attack kill chain: the node.js process is hijacked, and then converses with a remote malware server on port 2345 to download and execute a malicious script (exfil.py), which is eventually executed and starts an interaction with a command-and-control server on port 4444 to exfiltrate sensitive information from the customer database on port 3000.

This example showcases the advantages of applying flow analysis to system telemetry. \pname{} provides visibility within host environments by exposing relationships between containers, processes, files, and network endpoints as events (single operations) and flows (volumetric operations). For example, when the node.js process clones and execs into the new process, these tasks are recorded as process events (PE), and when a process communicates with a network endpoint or writes a file, these interactions are captured and summarized using compact file (FF) and network (NF) flows. The result is a graph-like data structure that enables precise reasoning and fast retrieval of security-relevant information, lifting system calls into a higher-level, compact representation that enables defense automation, scalable analytics, and forensics.  
\begin{figure*}[t]
     \vspace{-6pt}
     \centering
     \includegraphics[width=0.9\textwidth]{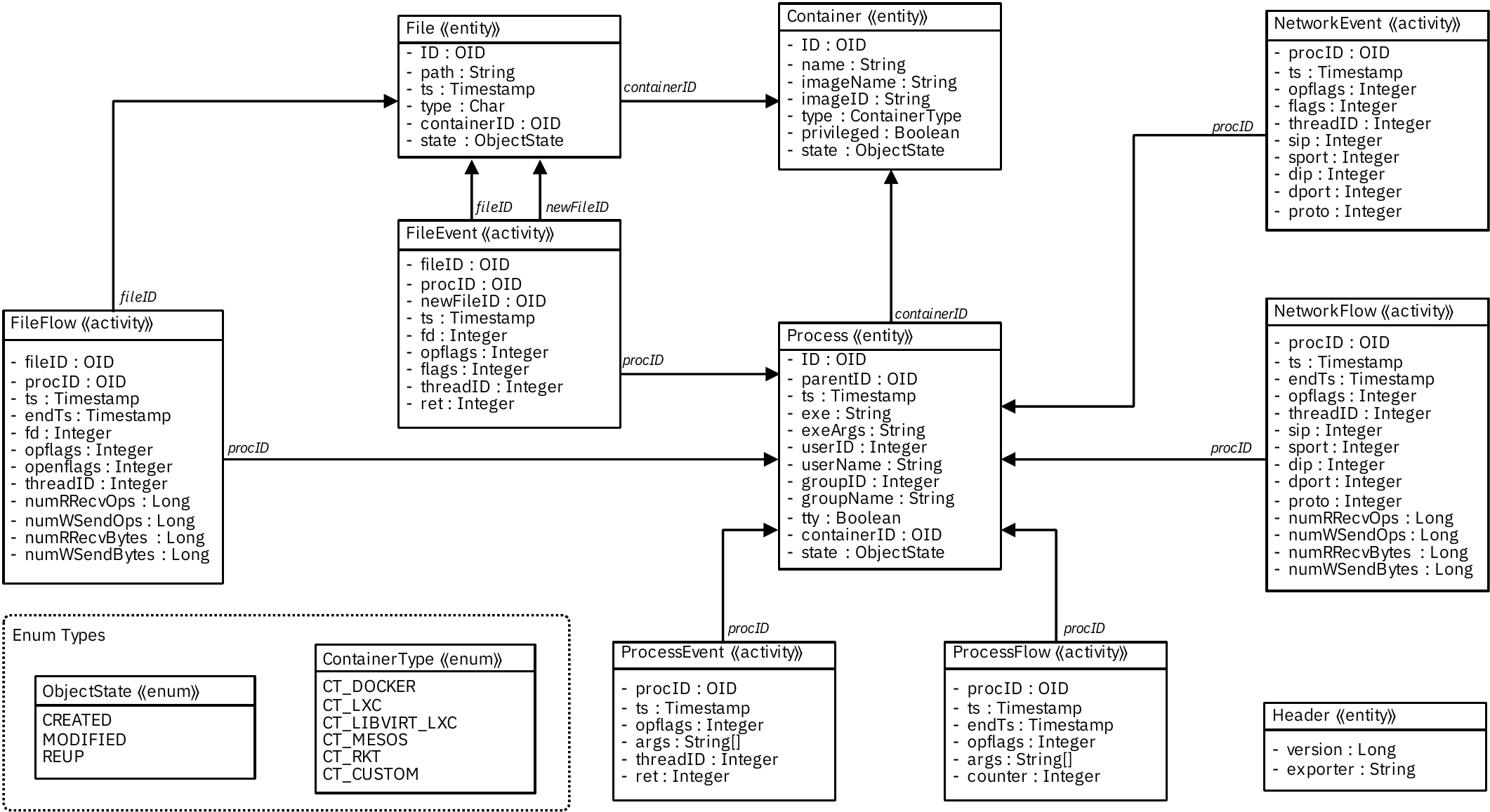}
     \caption{Overview of \pname{}'s object-relational view.}
     \label{fig:sysflowObj}
     \vspace{-6pt}
 \end{figure*}

\section{\pname{}}
\label{s:sysflow}

\pname{} is an open system telemetry format that lifts system call events into behaviors that describe how processes interact with system resources. 
Its design emphasizes security visibility, with increased importance on linking process behaviors for threat hunting, reducing data footprints for forensic storage, and providing a rich list of attributes to identify malicious techniques, tactics, and procedures (TTPs).  To support these goals, \pname{} aggregates related system calls into compact summarizations that are time-bound and contain a vast number of carefully selected  attributes and statistics, ideal for a wide variety of analytics and forensics. The format's object-relational abstraction 
inherently encodes the graph-like structure of a process interacting with its environment, which is ideal for gaining contextual information about attacks, and supporting threat hunting. Figure~\ref{fig:sysflowObj} shows an overview of \pname{}'s main objects and relations, which are categorized as \textit{entities}, \textit{events}, and \textit{flows}.

\subsection{Entities}
\label{ss:entities}
Entities represent components or resources that are monitored on a system. The current specification supports four types of entities: \textit{header}, \textit{container}, \textit{process}, and \textit{file}. The header object contains runtime and meta-data information about the host system being monitored, including hostname, distribution, kernel version, and \pname{}'s specification version.  

Entities (except the header) contain a timestamp, an object ID and a state along with other entity-specific attributes (e.g., container name, exe, pid). The timestamp attribute is used to indicate the time at which the entity was exported to the captured trace. Each entity has an object ID to allow events, flows and other entities to reference each other without having duplicate information stored in each record and to support rapid process graph building on \pname{} data streams. Object IDs are not required to be globally unique across space and time. In fact, the only requirement for uniqueness is that no two objects managed by a \pname{} exporter can have the same ID simultaneously. Entities are always written to the trace file before any events, and flows associated with them are exported. Since entities are exported first, each event, and flow is matched with the entity (with the same ID) that is closest to it in the file. 

The container object represents a container-based workload, such as Docker or LXC, and is important to support the telemetry collection of modern cloud environments based on technologies, such as Kubernetes~\cite{k8s:2019}, Docker Swarm~\cite{dockerswarm:2019}, and OpenShift~\cite{openshift:2019}. It contains information about its image, name, type and ID. Central to the format is the process object, which contains a reference ID to the container object, as well as its parent process object ID. This greatly facilitates the tasks of relating processes to their containers and reconstructing the entire process tree. 

\subsection{Events \& Flows}
\label{ss:evflow}
Entity behaviors are modeled as \textit{events} or \textit{flows}.  These behaviors typically involve a process interacting with some resource (e.g., file, network, or another process) and are created from system call tracing. Events represent important \textit{individual behaviors} of an entity that are broken out on their own due to their importance, their rarity, or because maintaining operation order is important. An example of an event would be a process \code{clone} or \code{exec}, or the deletion or renaming of a file. By contrast, a flow represents a \textit{volumetric aggregation} of multiple events that naturally fit together to describe a particular behavior. For example, we can model the network interactions of a process and a remote host as a bidirectional flow that is composed of several events, including \code{connect}, \code{read}, \code{write}, and \code{close}. During extensive experimentation with raw system call data collection, we observed that file reads/writes, network sends/receives, and process threads creations are the main contributors to large trace sizes in conventional system telemetry approaches. This motivated our key insight that \textit{flows} can significantly reduce data footprints without loss of visibility. Flows are summarized by thread ID, meaning that a flow started by one process or thread and passed to another, will generate two flow records.

All events and flows contain a process object ID, a timestamp, a thread ID, and a set of operation flags (op flags). The process object ID represents the process entity on which the event or flow occurred, while the thread ID indicates which process thread was associated with the event. Behaviors are encoded in the op flags and represented in a single bitmap which enables multiple actions to be easily combined. The difference between  events and flows is that an event will have a single bit active, while a flow could have several. Flows represent a number of events occurring over a period of time, and thus have a set of operations (multiple bits set in the op flags), a start and an end time. One can determine the operations in the flow by decoding the op flags. A flow also contains a set of counters to record the number of operations over the duration of the flow.  

A flow is started by any supported flow operation and is exported on (1)~an \code{exit} event of its owning process, (2)~a \code{close} event signifying the end of a connection or file interaction, or (3)~a preconfigured timeout. After a long-running flow is exported, its counters and flags are reset; if no flow activity is observed during the timeout period, that flow is not exported.

\inlinedsectionit{Operations} Table \ref{t:opTable} shows the list of operations supported by \pname{}. It comprises 31 operations that map to one or more system calls. Note that \pname{} does not seek to provide coverage of every system call supported by the kernel---it would be far too costly to store and analyze. Rather, the goal is to provide a low-noise, \emph{semantics-preserving} telemetry source that records how processes affect their environment. These types of behaviors are extremely important for security and performance analyses including identifying cyber attack tactics, techniques and procedures, and are highly compressible.  

The underlying operational abstraction reveals that event and flow objects can represent process (control), file (access) and network (interaction) activity. We detail these next.

\begin{table}[t]
\centering
\scriptsize
\caption{Operations breakdown by event and flow type.}
\begin{tabular}{@{}l@{\hskip.15cm}p{7cm}@{}}
\toprule
\textbf{Event/Flow} & \textbf{Supported Operations} \\ 
\midrule
  Process Event & clone (process), exec, exit (process), setuid, setgid \\ 
\cmidrule{2-2}
  Network Event & bind, listen \\
\cmidrule{2-2}
  File Event & mkdir, rmdir, unlink, symlink, link, rename, chmod, chown, (u)mount \\
\cmidrule{2-2}
  Process Flow & clone (thread), exit (thread) \\
\cmidrule{2-2}
  Network Flow & accept, connect, send, recv, shutdown, close \\
\cmidrule{2-2}
  File Flow & open, read, write, setns, mmap, close \\ 
\bottomrule
\end{tabular}
\label{t:opTable}
\end{table}



\begin{table*}[t]
\vspace{-6pt}
\scriptsize
\centering
\caption{Simplified \pname{} records (PE=Process Event, FF=File Flow, FE=File Event, NF=Network Flow).}
\begin{tabular}{llllllllllll}
\toprule
\textbf{\#} & \textbf{Type} & \textbf{Process} & \textbf{PPID} & \textbf{PID} & \textbf{Op Flags} & \textbf{Start Time} & \textbf{End Time} & \textbf{Resource} & \textbf{Reads} & \textbf{Writes} & \textbf{Cont ID} \\
\midrule
1 & PE & node app.js  & 1887 & 21847 & EXEC & 4/10T16:47 & & & : & : & node-js \\
2 & FF & node app.js  & 1887 & 21847 & O  R  C & 4/10T16:47 & 4/10T16:47 & /lib/gnu/libc.so & 1:832 & : & node-js \\
3 & FE & node app.js  & 1887 & 21847 & MKDIR & 4/10T16:47 &  & /tmp/log  & : & : & node-js \\
4 & FF & node app.js  & 1887 & 21847 & O  W   & 4/10T16:47 & 4/10T16:48 & /tmp/log/app.log & : & 100:8000  & node-js \\
\rowcolor[gray]{.85} 5 & NF & node app.js  & 1887 & 21847 & A SR C & 4/10T16:48 & 4/10T16:49  & \scriptsize{<IP\textsubscript{attacker}>:3522 -- 172.30.10.2:443}  & 1:80 & 2:980 & node-js \\
\rowcolor[gray]{.85} 6 & NF & node app.js  & 1887 & 21847 & C SR C & 4/10T16:48 & 4/10T16:49  & \scriptsize{172.30.10.2:8353 -- <IP\textsubscript{Malware}>:2345}  & 3:4355 & 1:94 & node-js \\
\rowcolor[gray]{.85} 7 & FF & node app.js  & 1887 & 21847 & O  W  C & 4/10T16:48 &
4/10T16:48 & /tmp/exfil.py & : & 6:4250  & node-js \\
\rowcolor[gray]{.85} 8 & PE & /tmp/exfil.py  & 21847 & 21849 & EXEC & 4/10T16:48 & & & : & : & node-js \\
\rowcolor[gray]{.85} 9 & PE & apt install pip  & 21849 & 21851 & EXEC & 4/10T16:48 & & & : & : & node-js \\
\rowcolor[gray]{.85} 10 & PE & apt install pip  & 21849 & 21851 & EXIT & 4/10T16:48 & & & : & : & node-js \\
\rowcolor[gray]{.85} 11 & NF & /tmp/exfil.py  & 21847 & 21849 & C SR C & 4/10T16:48 & 4/10T16:48  & \scriptsize{172.30.10.2:8356 -- <IP\textsubscript{Storage}>:3000}  & 2:165 & 1:34 & node-js \\
\rowcolor[gray]{.85} 12 & NF & /tmp/exfil.py  & 21847 & 21849 & C SR C & 4/10T16:48 & 4/10T16:48  & \scriptsize{172.30.10.2:8357 -- <IP\textsubscript{C\&C}>:4444}  & 1:46 & 2:188 & node-js \\
\rowcolor[gray]{.85} 13 & PE & /tmp/exfil.py  & 21847 & 21849 & EXIT & 4/10T16:48 & & & : & : & node-js \\
14 & FF & node app.js  & 1887 & 21847 &   W   & 4/10T16:48 &
4/10T16:49 & /tmp/log/app.log & : & 100:8000  & node-js \\
15 & FF & node app.js  & 1887 & 21847 &   W C  & 4/10T16:49 & 4/10T16:49 & /tmp/log/app.log & : & 50:4000 &  node-js \\
16 & PE & node app.js  & 1887 & 21847 & EXIT & 4/10T16:50 & & & : & : & node-js \\
\bottomrule
\label{t:exFlows}
\end{tabular}
\vspace{-7pt}
\end{table*}

\inlinedsectionit{{Process Activity}} Process events and flows represent behaviors that modify a process' control flow.   
The difference between a process event and a flow, is that the process event is emitted every time there is a modification to the process (e.g., change user ID), a process \code{exit} is observed, or a new process is spawned. By contrast, a process flow keeps a summary of new threads created and destroyed over a time period. Process flows are created on \code{clone} events, and only emitted when the flow expires (after a timeout), or when the process exits.

Table~\ref{t:exFlows} shows a sample of the \pname{} trace discussed in \S\ref{ss:motivation}, which describes the launching of a node.js server, acceptance of an incoming HTTPS request, dropping and execution of a malicious script followed by the termination of both applications running in container \code{node-js}. Rows highlighted in gray correspond to suspicious behavior. Records \#1 and \#8 captures the process events corresponding to the execution of process \code{node} and suspicious script \code{exfil.py}, respectively. Moreover, record \#9 shows the execution of package installer \code{apt}, whose parent PID (PPID) corresponds to the suspicious script's PID. Finally, records \#10, \#13 and \#16 signal that the processes have terminated. Note that the trace only displays a subset of the record attributes due to space constraints.   

\inlinedsectionit{{File Activity}} File events and flows represent process interactions with files, pipes, UNIX sockets, and devices on a system.  File events focus on self-contained system calls that create, delete, or rename files and directories.
Conversely, file flows summarize all the operations executed on the same file handle.  A file flow will typically begin on a file \texttt{open}, summarize operations on that handle, and then expire on a file \texttt{close}, or after a timeout, after which flow counters are reset. File flows keep track of the number of \code{read} and \code{write} operations on a file, as well as the number of bytes read and written.   

To illustrate, record \#2 denotes a file flow in which the \code{node} process reads 832 bytes from the \code{libc.so} shared library in one \code{read} operation ({1:832}). The \code{O R C} operation flags indicate that \code{node} \underline{o}pened, \underline{r}ead, and \underline{c}losed the shared object. The process then creates a \code{/tmp/log} directory as captured by the file event in record \#3.  
The process continues by writing to \code{app.log} over a 2-minute period, recorded as three file flows (\#4, \#14--15). The first flow shows when the file is opened and 8000 bytes are written using 100 \code{write} operations. The flow expires, and a \emph{continuation} flow is created, which records the next 8000 bytes written. Finally, the second flow expires, and a third flow is exported with the remaining writes and the final \code{close} operation. In between writes to the log file, the node server is hijacked, and an \code{exfil.py} is written as a file flow (\#7) and then executed. 

\inlinedsectionit{{Network Activity}}  Network events and flows are analogous to their file counterparts but work on sockets rather than files. A network flow records all interactions of a process with a local or remote IP address containing the same 5-tuple (IP addresses, ports, proto) as Cisco NetFlow. Network flows also contain counters for the number of bytes sent and received similar to bidirectional NetFlows.

Network flows differ from NetFlows in that they operate at the transport layer, by monitoring  and summarizing network-related system calls such as \code{accept}, \code{connect}, \code{send}, or \code{recv}. It requires the application to explicitly interact with a remote/local endpoint in order to be generated. By contrast, NetFlow operates at the Internet/Network (or packet) layer. This means that NetFlow has the concept of packets and TCP Flags (which network flows do not), and can record traffic from remote hosts that is ignored by the local system. 

While network flows do not record certain remote traffic patterns like scanning activity, they record active process connections and allow network traffic to be tied to process information as well as file information, providing a more granular and comprehensive data source for security and performance analytics. Record \#5 is a network flow representing the attacker's connection to the \code{node} server. The server \underline{a}ccepts the connection from the attacker, \underline{s}ends and \underline{r}eceives data, and finally \underline{c}loses the connection, as denoted by the operation flags. After \code{exfil.py} is written and executed, it connects to a storage service on port 3000 (\#11) and begins data exfiltration to a command-and-control server (\#12) on port 4444.

\subsection{Extended Attributes}
\label{ss:ext}
Certain attributes, such as domain names and file hashes, are not obtainable through system calls but can be gathered through other mechanisms to enrich \pname{} records. These are captured as extended attributes and are not depicted in Figure~\ref{fig:sysflowObj}.

\section{Design \& Implementation}
\label{s:design}
Our implementation comprises a collection stack and a data processing pipeline for \pname{} built atop the Linux operating system. We also describe a cloud-native telemetry architecture that can store and analyze \pname{} data at scale. Other telemetry stacks and operating systems such as Windows can also benefit from utilizing \pname{}'s open schema, but a more detailed discussion is beyond the scope of this paper. 

\subsection{Data Collection}
\label{s:probe}
We built our collector using Sysdig's core system telemetry libraries (v0.26.7)~\cite{Sysdig:2019}.
Sysdig uses a kernel module and tracepoints to dump system call information into a ring buffer, which is in turn processed by Sysdig's core libraries\footnote{Sysdig also supports eBPF probing since version 0.26.4.}. We chose Sysdig because it is an open source project with strong community support, has a powerful API that supports container monitoring natively, and provides all the attributes \pname{} requires. Other telemetry options such as Linux Audit do not support container as first class objects. Our implementation is \textasciitilde4k lines of \Cplusplus{} code. The telemetry stream is serialized using Apache Avro~\cite{Avro:2012}
(see \S\ref{ss:benchmarks} for an empirical discussion of serialization formats).

\subsection{Cloud-native Telemetry Pipeline}
\label{s:language}

\begin{figure}[t]
     \centering
     \includegraphics[width=0.95\hsize]{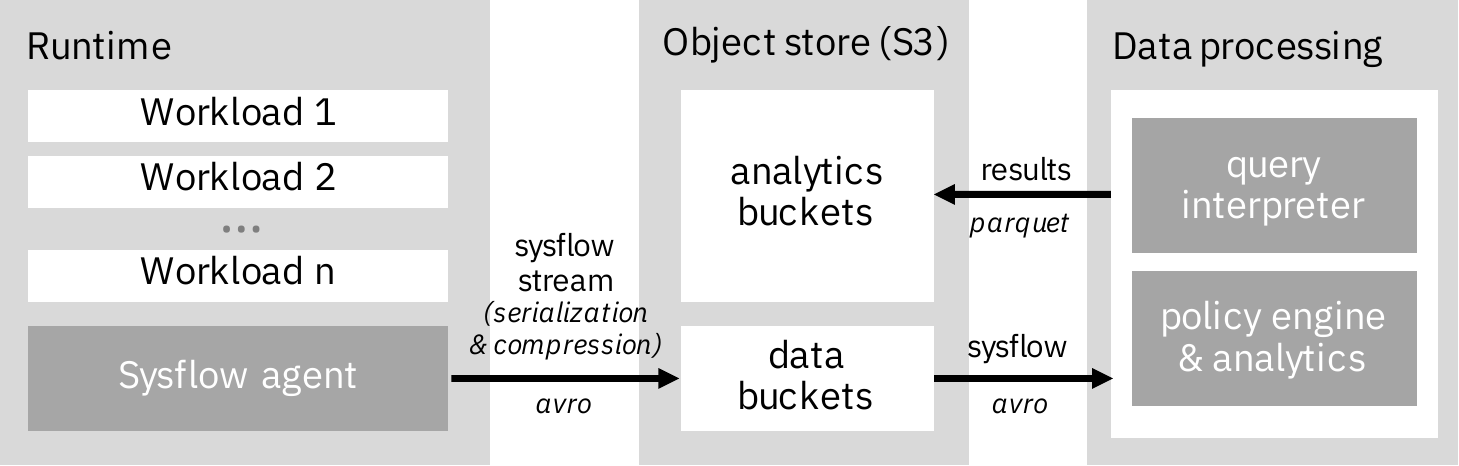}
     \caption{\pname{} data processing pipeline.}
     \label{fig:pipeline}
     \vspace{-6pt}
 \end{figure}

To enable efficient processing and analysis of \pname{} data, we implemented a cloud-native telemetry pipeline for large-scale system analytics. The pipeline, shown in Figure~\ref{fig:pipeline}, deploys and replicates the \pname{} agent (collector and exporter) to every Kubernetes~\cite{Kube:2019} node (daemonsets) of monitored cloud clusters, using Helm charts~\cite{Helm:2019} for deployment automation. The agent streams \pname{} objects to files on each node, which are subsequently pushed, at predefined intervals, to buckets in a distributed S3-compliant cloud object store
Once exported, both microservice- and Spark-based
analytics are employed to continuously process telemetry data. 

\inlinedsectionit{Querying \& Analytics} Data introspection is made possible via a set of programmatic and declarative APIs offered to telemetry consumers via a simple, yet powerful processing language that supports both alerting and enrichment of telemetry records.
We used these APIs to implement language runtimes and a declarative policy engine for both Spark (Java) and Python, and are exploring scalable machine learning approaches to detect security incidents in cloud environments. Note that since objects are serialized Avro, we can natively support analytics in any major programming language, as well as big data platforms.     
 
\inlinedsectionit{Processing Language} 
We designed a domain-specific language along with its distributed runtime for processing \pname{} traces at scale. The language is declarative and embeds a simple predicate logic extended with operators, lists and macro definitions. For explanatory precision, a simplified version of the language is shown in Figure~\ref{fig:syntax}.

Policies $\policy$ are sets of lists, macros, and rules. A $\flist$ is an array of values (e.g., names of common package installers). A $\fmacro$ is a named condition, denoted $\cond$. A $\frule$ defines an action, which can be to $\tmatch$ a stream of \pname{} records with $\cond$ (and optionally $\tshow$ the  records $\conj{\sysflow}$ satisfying the constraints defined in $\cond$), or to $\ttag$ the list of matched records with label set $\conj{\val}$. Conditions are defined as logical expressions predicated over primitive \pname{} attributes (e.g., $sf.proc.exe$, $sf.file.path$) and attribute extensions defined in the language. For example, one useful attribute extension is the ability to predicate over ancestry chains, like in the rule: 
\begin{center}
{\small$\tmatch$} {\small$sf.proc.achain(2)$} {\small$\tin$} {\small\textit{shell\_binaries}}
\end{center}
which matches all records with a shell as grandparent.  


\begin{figure}[t]
\vspace{-5pt}
\scriptsize
\begin{mdframed}[style=frame,rightline=false,leftline=false,innerleftmargin=0pt,innertopmargin=-5pt,innerbottommargin=-5pt]
\begin{align*}
%
%
&\grsymbol{policies} & \policy &\production \mult{(\flist\choice\fmacro\choice\frule)}\\
&\grsymbol{lists} & \flist &\production \var\kern1pt{\defn}\kern1pt{\conj{\val}} \\	
&\grsymbol{macros} & \fmacro &\production \var\kern1pt{\defn}\kern1pt\cond  \\	
&\grsymbol{rules} & \frule &\production  \tmatch~\cond~[~\tshow~\conj{\sysflow}~] \\
                            &&&\choice \ttag~\cond~\twith~\conj{\val}  \\
&\grsymbol{conditions} & \cond &\production \expr~\multopt{(\bor~\expr)} \\
&\grsymbol{~} & \expr &\production \term~\multopt{(\band~\term)} \\
&\grsymbol{~} & \term &\production \var \choice \opnegt\term \choice \op~\unop \choice \op_1~\binop~\op_2  \\
&\grsymbol{unary ops} & \unop &\production \texists \\
&\grsymbol{binary ops} & \binop &\production  \tin \choice \tpmatch \\ 
				            &&&\choice  \tcontains \choice \tstartswith \choice \ldots \\
							&&&\choice \text{typical comparison operators} \\
&\grsymbol{atoms} & \op &\production \val \choice \var \choice \sysflow \\
&\grsymbol{variables} & \var & \\
&\grsymbol{values} & \val &\production \text{values of the underlying language} \\
&\grsymbol{records} & \sysflow &\production \text{sysflow record attributes} \\
\end{align*}
\end{mdframed}
\caption{\pname{} simplified processing language syntax.}
\label{fig:syntax}
\vspace{-10pt}
\end{figure}

\begin{table*}[t]
\vspace{-6pt}
\scriptsize
\centering
\caption{MITRE ATT\&CK TTPs that can be observed in \pname{} streams and expressed as behavioral policies; P=Process, F=File, PE=Process Event, PF=Process Flow, FE=File Event, FF=File Flow, NE=Network Event, NF=Network Flow.}
\begin{tabular}{l @{\outersep} c @{\outersep} m{13.5cm}}
\toprule
\textbf{Type} & \textbf{Behavior} & \textbf{Expressible TTPs} \\
\midrule
PE & \raisebox{-0.4\totalheight}{\includegraphics[width=0.061\textwidth]{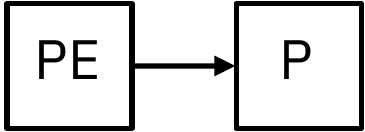}} & process injection, process modification, process discovery, hijack execution flow, event trigger execution, 
command and scripting interpreter abuse, native API exploit, scheduled task/job, software deployment tools, account manipulation, \tcode{modprobe} invocation, application instability/crash detection, spikes in application activity, \tcode{sudo} execution, \tcode{kill} signals to security applications, compiler executions, package installation, history tampering (\tcode{HISTCONTROL}), library injection (\tcode{LD\_PRELOAD}), regexes in grep commands, local/logged user listing, filesystem scanning, account discovery, group execution, network discovery, remote file copy execution, compression binaries execution, screen capturing execution, system shutdown/reboots \\
\cmidrule{3-3}
PF & \raisebox{-0.4\totalheight}{\includegraphics[width=0.061\textwidth]{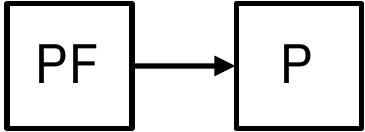}} & spikes in application activity, thread instability/trashing behavior \\
\cmidrule{3-3}
FE & \raisebox{-0.4\totalheight}{\includegraphics[width=0.1\textwidth]{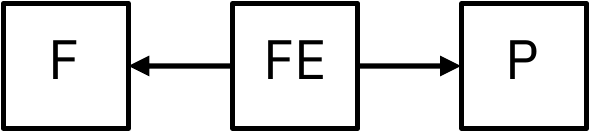}} & filesystem access control modification, \tcode{mknod} execution, file deletions, file metadata modification, hidden directory creation, spikes in application activity, \tcode{/var/log} deletion \\
\cmidrule{3-3}
FF & \raisebox{-0.4\totalheight}{\includegraphics[width=0.1\textwidth]{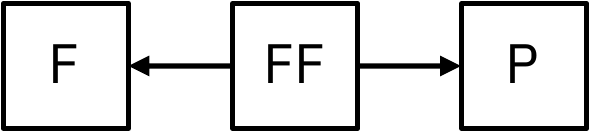}} & account discovery, file download, protected file writes, configuration file modification, scheduled files/crontab modification, file name injection, bash profile modification, \tcode{/etc/passwd} modification, hidden file creation, read from \tcode{.ko} files, installation of new classes/libraries, new classes/libraries loaded in application startup, system init process modification, \tcode{sudoers} file modification, executable file modification, bash history modification, root certificate modification, credentials scanning, unusual access to cached password data \\
\cmidrule{3-3}
NE & \raisebox{-0.4\totalheight}{\includegraphics[width=0.061\textwidth]{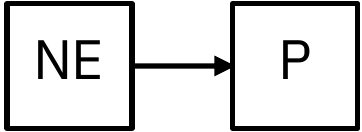}} & unusual port bind and listen operations \\
\cmidrule{3-3}
NF & \raisebox{-0.4\totalheight}{\includegraphics[width=0.061\textwidth]{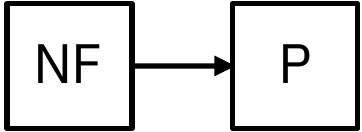}} & network sniffing, communication with new network device, interactions with known malicious site, strange SSH login attempts, failed SSH logins, beaconing/periodic patterns in network traffic, suspicious correlation between process and network activity, unusual access patterns of shared drives or remote repositories, unusual ports, spike in network activity\\
\bottomrule
\label{t:mitre}
\end{tabular}
\vspace{-12pt}
\end{table*}

\subsection{Automated Attack Detection}
In this section, we return to the remote attack presented in \S\ref{ss:motivation}. and investigate how the policy language and analytics pipeline can be used to detect such an attack and perform threat hunting. \pname{} collectors monitor the container workloads in the targeted environment and push \pname{} records into the analytics pipeline described in \S\ref{s:language}.  The pipeline is equipped with a distributed policy engine based on \pname{}'s processing language.
Any container actions that violate a security policy trigger an alert.

\inlinedsectionit{Security Policies}
A typical policy is that containers requiring an update must be rebuilt and redeployed. Such policy must therefore prohibit someone from executing a system package installer (e.g., \texttt{yum}, \texttt{apt}) inside a running container. This policy can be defined as shown in Rule~\ref{lst:rule:package-management}, which alerts on the execution of any software management tool.

\begin{lstlisting}[language=bash, caption=Detection of package manager execution., label=lst:rule:package-management,  basicstyle=\normalfont\small, aboveskip=6pt, belowskip=10pt, numbersep=0pt, resetmargins= true]
 @\tmatch~{\small$sf.proc.exe$}~\tpmatch~{\footnotesize(apt, yum, dnf, \ldots)}@
\end{lstlisting}

Rule~\ref{lst:rule:package-management} alerts the bootstrap phase of the attack, where \texttt{apt} (see Table~\ref{t:exFlows}) is invoked for the preparation of the Python runtime. The policy engine implements rules that inspect container integrity, access control operations, known attack TTPs, and application-specific misbehavior during container lifecycle.

\inlinedsectionit{Policy Expressiveness} \pname{} can express security-relevant behaviors succinctly and without the need to record raw system call information. Table~\ref{t:mitre} summarizes our analysis of the MITRE's ATT\&CK~\cite{Mitre:2019}, in which we show how different \pname{} components can be used to characterize the vast majority of tactics, techniques, and procedures (TTPs) described in the framework (e.g., Rule~\ref{lst:rule:tag}). \pname{}'s graph-relational structure captures system events in the context of other surrounding events, which enables the characterization of behavioral patterns and provenance chains. Our flow-based abstractions occlude inter-arrival times and individual system call granularity for certain operations; however, we find this to be an acceptable trade-off for most analyses. Moreover, in circumstances where more granularity is required, analysis can be performed on streamed system call data, with the raw telemetry stored in \pname{} format for evidence preservation and forensics. 

\begin{lstlisting}[language=bash, caption=Tagging of MITRE's T1087 (\emph{account discovery})., label=lst:rule:tag,  basicstyle=\normalfont\small, aboveskip=6pt, belowskip=10pt, numbersep=0pt, resetmargins= true]
 @\ttag~{\small$sf.file.path$}~\tin~{\footnotesize(/etc/passwd, /etc/shadow, \ldots)}~\twith~{\footnotesize[T1087]}@
\end{lstlisting}

\inlinedsectionit[0pt]{Cyber Threat Hunting}
In security operation centers, analysts dequeue alerts from automated detectors, triage them, search for their context, and reason about them for the bigger picture of multi-phase attack campaigns. The system telemetry provided by \pname{}  enables comprehensive context completion, provenance tracking, and reasoning to hunt threats. For each alert, we implemented rules to automatically extract context from related entities, such as \emph{call chains} (Rule~\ref{lst:rule:ancestry-chain}) and \emph{loaded files}.

\begin{lstlisting}[language=bash, caption=Show ancestry chain of a process., label=lst:rule:ancestry-chain,  basicstyle=\normalfont\small, aboveskip=6pt, belowskip=10pt, numbersep=0pt, resetmargins= true]
 @\tmatch~{\small$sf.proc.exe$}~\tcontains~{\footnotesize{exfil.py}}~\tshow~{\small$sf.proc.achain$}@ 
\end{lstlisting}

More advanced capabilities can be coded by threat hunters to explore the neighborhood of alerts, mine connections, and track data provenance. In our pipeline, analysts can use a SQL interface as the query language, and the queries are translated into the \pname{} processing language. Starting from the prohibited execution of \texttt{apt} and the beaconing alerts, the analyst quickly locates the \texttt{node.js} worker, which is on both the call chain of the Python script that exfiltrates data and the call chain of the \texttt{apt} process that precedes the script execution. The attack is fully revealed when activities of processes on the call chain are displayed, which includes environment setup with \texttt{pip}, object store search, and data retrieval from the storage service.

\section{Evaluation}
\label{s:evaluation}
To demonstrate the efficacy of \pname{}, we monitored popular cloud services under strenuous benchmark conditions and measured their throughput, resource utilization, and performance impact on the workloads. We also compared our collector implementation with two mainstream telemetry frameworks: go-audit~\cite{go-audit:2016} and Sysdig~\cite{Sysdig:2019}. In summary, our collector generates \textit{orders of magnitude less data} than the syscall-granular alternatives, especially when monitoring I/O-heavy applications, such as databases and object stores. Furthermore, our collector does not affect the performance of monitored workloads, while using less than one vCPU under the heaviest workload profiles. 

\subsection{Experimental Setup}
\label{ss:expsetup}
Experiments were conducted on 4 virtual machines (VMs) hosted on a popular cloud platform, each containing 48 2.0GHz cores, 192 GB of RAM, 2 TB of disk space, and running Ubuntu 18.04. Two VMs were used as load testers while the other two hosted the targeted workloads; workloads (except Hadoop) were deployed as docker containers.  

To cope with the magnitude of the experiments, we built an automated evaluation framework and testing harness based on Ansible~\cite{ansible:2019} playbooks. Ansible allowed us to deploy both workload and monitoring containers on the desired machines while kicking off load tests and gathering performance statistics. \texttt{psrecord}~\cite{psrecord:2013} was used to measure CPU and memory usage on all workloads and collectors. 

Sysdig was configured to use zlib block compression on its data, while \pname{} used the deflate codec for block compression. Furthermore, \pname{} was configured to timeout and export long-running file and network flows every 30 seconds.
All three telemetry sources filter for the same set of system calls. Finally, go-audit does not support filtering by container natively. As a result, we ran containers and processes under a special user ID, and filtered system calls on that user.  

\subsection{Performance Benchmarks}
\label{ss:benchmarks}

\begin{table}[t]
    \centering
    \footnotesize
    \setlength{\tabcolsep}{3pt}
    \caption{Benchmarks used in the evaluation.}
    \begin{tabular}{llp{4.2cm}}
        \toprule
        \textbf{Workload} & \textbf{Benchmark}  & \textbf{Settings} \\
        \midrule
         httpd\_prefork &  ab~\cite{ab:2019} & $c$: 1,5,10,25,50 $t$: 60s $n$: 100M  \\
         httpd\_worker &  ab~\cite{ab:2019} & $c$: 1,5,10,25,50 $t$: 60s $n$: 100M  \\
         httpd\_event &  ab~\cite{ab:2019} & $c$: 1,5,10,25,50 $t$: 60s $n$: 100M  \\
         minio & Wasabi S3 BM~\cite{wasabi:2019} & $threads$: 1, 5, 10 $dur$: 15m $file~size$: 5, 64, 256 (MB) \\
         mysql & TPC-H HDB~\cite{hammerdb:2019} & default \\
         postgres & TPC-C HDB~\cite{hammerdb:2019} & default \\
         redis & TPC-C HDB~\cite{hammerdb:2019} & default \\         
         hadoop & HiBench~\cite{hibench:2019} & $scale$: large, $tests$: wordcount, sql scan, pagerank  \\
         matrix\_mult & m-thread bench~\cite{maigre:2016} & $benchmark$: 3, $matrix~size$: 5000x5000, 1 thread per row \\
         \bottomrule
    \end{tabular}
    \label{t:benchmarks}
\end{table}

Table \ref{t:benchmarks} shows the list of workloads and benchmarks used in the evaluation.  We chose these applications for their wide popularity in cloud-based deployments and plurality in terms of roles and software architecture (e.g.,  process intensive, thread intensive, file intensive, network intensive), enabling the creation of very diverse telemetry footprints. 

We also selected a set of standard industry benchmarks that tax workloads with higher than normal traffic patterns to demonstrate the capabilities of the collectors under intense pressure. Benchmarks include data processing (TPC-C), decision support (TPC-H), web traffic generation (Apache Benchmark),  S3 object creation, retrieval and deletion, multi-threaded matrix multiplication, as well as big data analytics.   

\begin{figure}[t]
    \centering    
    \includegraphics[width=.95\linewidth]{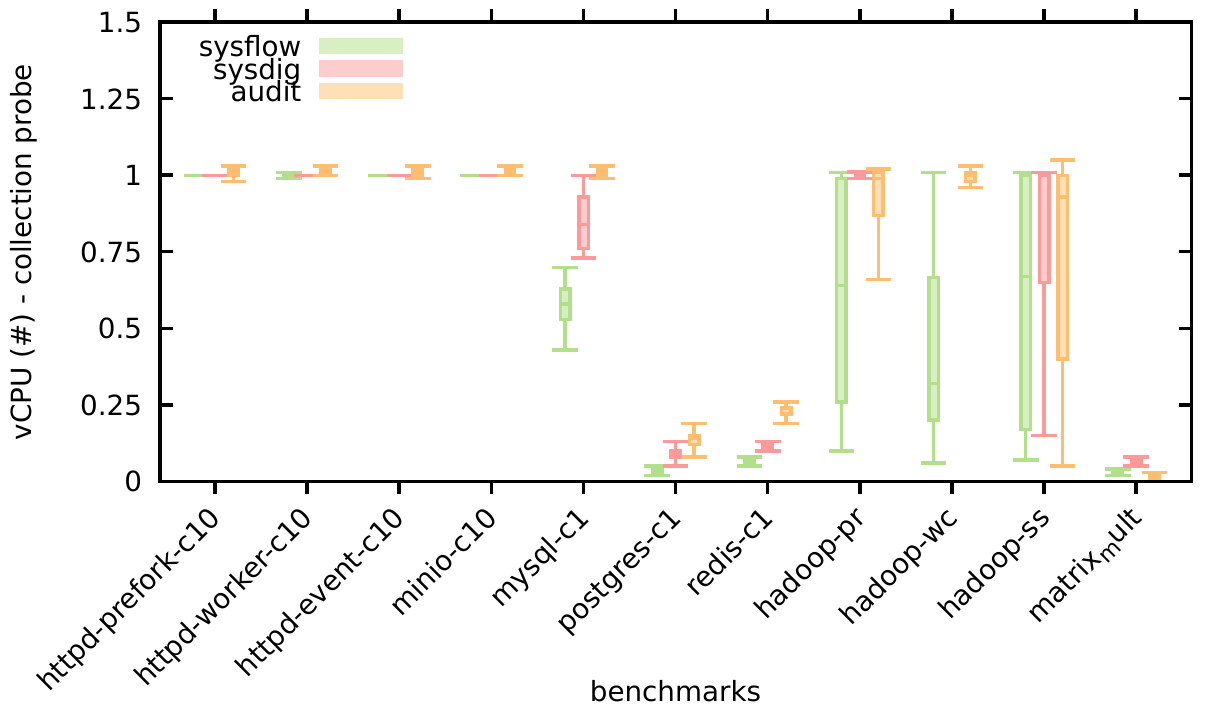}
    \caption{CPU utilization. The benchmark name suffixes denote the number of concurrent clients $c$ for that benchmark.}
    \label{fig:mem_cpu}
\end{figure}

\inlinedsectionit{Collector CPU and Memory Utilization}
We measured the CPU and memory utilization statistics for all the collectors during the benchmarks. Figure~\ref{fig:mem_cpu}\footnote{Note that a subset of our tests are shown due to space constraints.} shows that the heavier the workload, the more CPU the collector uses, and all collectors use around the same amount of CPU for CPU-intensive benchmarks, such as the Apache Benchmark. Also, given that \pname{} is built on top of Sysdig's system call probe, the two collectors perform similarly in terms of CPU and memory usage, indicating that flow aggregation does not increase resource usage significantly. In fact,  \pname{} outperforms the other collectors on the database benchmarks because its output is reduced by several orders of magnitude more than the other sources. As a result, our collector is not doing block compression or intensive I/O like the other collectors. In general, CPU usage varies wildly for Hadoop due to the stop and start nature of its jobs.  None of the collectors used more than 100MB of memory.

\inlinedsectionit{Workload CPU and Memory Utilization}
The CPU utilization for the tested workloads when under monitoring and without monitoring (baseline) indicates that the collectors have negligible performance impact on the workloads. Similarly, the collectors did not affect workload memory usage. This is expected as the collectors passively monitor for system calls via kernel probes, unlike system call interposition techniques.


\begin{table}[t]
\centering
\setlength\tabcolsep{6pt}
\footnotesize
\caption{Trace sizes (MB) and parenthesized object counts ($\times10^3$) across benchmarks.}
\begin{tabular}{l@{\hskip.5cm} r@{\innersep}l @{\outersepsmall} r@{\innersep}l @{\outersepsmall} r@{\innersep}l}
\toprule
\textbf{Benchmark} & \multicolumn{2}{c}{\textbf{SysFlow}} & \multicolumn{2}{c}{\textbf{Sysdig}} & \multicolumn{2}{c}{\textbf{Go-Audit}}\\ 
\midrule
httpd\_prefork\_c10	& 11 &	\tbvar{818.59} & 62 & \tbvar{7,451.43} & 78 &	\tbvar{194.11} \\
httpd\_worker\_c10 & 7.9 &	\tbvar{629.31} & 58 &	\tbvar{6,399.76} & 80 &	\tbvar{193.28} \\
httpd\_event\_c10 & 7.8 &	\tbvar{642.35} & 53 &	\tbvar{5,990.95} & 75 & \tbvar{184.80} \\
mysql\_bs	& 0.1 & \tbvar{1.42} &   3735 & \tbvar{139,104.20} &  794 & \tbvar{1,888.74} \\
mysql\_c1	& 0.09 & \tbvar{0.19} &  2592 & \tbvar{156,002.41} &  327 & \tbvar{786.26} \\
postgres\_bs & 0.07 & \tbvar{1.53} &  7.8 & \tbvar{825.19} &   188 & \tbvar{400.69} \\
postgres\_c1 & 0.22 & \tbvar{7.08} &  169.2  & \tbvar{9,314.71} & 2000 & \tbvar{4,291.52} \\
redis\_bs &	 0.1 & \tbvar{6.04} &   160.8 & \tbvar{8,687.36} & 2500 & \tbvar{4,253.15} \\
redis\_c1 &	 0.15 & \tbvar{9.10} &  68 & \tbvar{15,240.27} &    4100 & \tbvar{8,309.04} \\
minio\_c10 & 19.5 & \tbvar{752.49} &  552 & \tbvar{26,241.50} &   966 & \tbvar{2,188.92} \\
hadoop\_pr & 16.4 & \tbvar{626.51} &  234 & \tbvar{10,945.46} &   1700 & \tbvar{2,469.77} \\
hadoop\_wc & 7.1 & \tbvar{259.92} &   45 & \tbvar{2,609.22} &    875 & \tbvar{1,239.34} \\
hadoop\_ss & 6.2 & \tbvar{225.86} &   42 & \tbvar{2,044.28} &    807 & \tbvar{1,128.15} \\
matrix\_mult & 0.008 & \tbvar{0.015} & 1.2 & \tbvar{30.1}  &  5.2  & \tbvar{11.49} \\
\bottomrule
\end{tabular}
\label{t:plot_sizes}
\end{table}

\inlinedsectionit{Semantic Compression}
Table~\ref{t:plot_sizes} shows the number of objects (records) and the file sizes generated by each of the collectors. At first sight, it appears that go-audit is the clear winner in generating the least number of objects among the collectors. However, this exposes a serious underlying problem---go-audit misses the vast majority of messages, especially in highly intensive benchmarks such as Apache Benchmark. Audit's internal buffer gets overwhelmed and starts spilling messages. 
This underscores the fact that audit is not the right tool for collecting system call events. Audit functions better as a rule-based runtime monitor whereby one can build rules to monitor specific processes or directories, limiting the scope of observed system calls. 
Furthermore, audit's lack of container support makes it a less than ideal choice for modern deployments.

When compared to \pname{}, Sysdig generates one order of magnitude more records at the low-end (httpd\_prefork) and six orders of magnitude more records at the high-end (databases). 
\pname{} performs much better on the database benchmarks because these databases are efficient, opening single file handles to each backend files, and leaving them open. Sysdig captures each read and write as individual records, whereas \pname{} creates one file flow per file, reexporting them every 30 seconds with updated statistics. 

\pname{} was closest in size to Sysdig on the Apache benchmarks. This is largely due to the fact that these benchmarks create thousands of quick network connections, which induce thousands of flows. Since there are not a lot of reads/writes per connection, the compression is smaller. 
An interesting aspect of the Apache web server under different multi-processing configurations is that it has a model where a root process or thread accepts a new connection, and then passes that connection handle off to another thread or process for processing. In some configurations, that connection will get passed to a third thread to be closed. In our prototype, we used the thread ID as an attribute to determine uniqueness in both network and file flows. This means that these Apache servers will generate up to three network flows per connection received. This was a conscious decision made for visibility; however, we could make the implementation configurable to only generate one flow when a connection is passed around, thus further reduce our output size. 

\pname{} provided over an order of magnitude compression on the Hadoop benchmarks compared to Sysdig and Audit. Hadoop is the most complex benchmark in that it creates tens of thousands of network and file flows as well as file and process events.  Indeed, the page rank benchmark used over 700 unique java processes, 34,000 unique files, and 38,000 file events including 2,400 \texttt{mkdir}s, and 31,000 \texttt{unlink}s.  Hadoop demonstrates \pname{}'s broad ability to compress telemetry data under a diverse set of system calls. 
The matrix multiplication benchmark was a thread intensive benchmark whereby, one thread was created per row of the resulting matrix.  \pname{} achieved high semantic compression by summarizing thread creation/destruction as a process flow.

\begin{table}[t]
\centering
\setlength\tabcolsep{6pt}
\footnotesize
\caption{\pname{}'s serialization format sizes (KB) computed from 1-min benchmark samples, with  overhead comparisons to native Avro format (parenthesized)}
\begin{tabular}{l@{\hskip1.2cm} r @{\outersep} r@{\innersep}l @{\outersep} r@{\innersep}l}
\toprule
\textbf{Benchmark} & \textbf{Avro} & \multicolumn{2}{c}{\textbf{JSON {\raise.5pt\hbox{\scriptsize(gzip)}}}} & \multicolumn{2}{c}{\textbf{Parquet}}\\ 
\midrule
httpd\_prefork	& 1,270 &	3,300 &	\tbvar{+160\%} & 1,334 & \tbvar{+5\%}\\
httpd\_worker &	1,874 &	4,900 &	\tbvar{+161\%} & 2,041 & \tbvar{+9\%}\\
httpd\_event & 2,320 &	5,800 &	\tbvar{+150\%} & 2,567 & \tbvar{+11\%}\\
mysql	& 75 & 118 & \tbvar{+57\%} & 85 & \tbvar{+13\%} \\
postgres & 61 & 87 & \tbvar{+43\%} & 68 & \tbvar{+11\%} \\
redis &	 52 & 93 & \tbvar{+79\%} & 59 & \tbvar{+13\%} \\
minio & 2,939 &	3,100 & \tbvar{+5\%} & 3,239 & \tbvar{+10\%} \\
hadoop & 11,119 & 18,483 & \tbvar{+66\%} & 9,682 & \tbvar{-13\%} \\
\bottomrule
\multicolumn{6}{l}{{\scriptsize$^*$Avro uses deflate codec with 80KB blocks; JSONs are compressed with gzip.}}\\
\end{tabular}
\label{t:json}
\end{table}

\inlinedsectionit{Object Serialization}
Table \ref{t:json} compares the size of \pname{} records serialized in both Avro, JSON (gzip compressed), and Parquet. Notice that even in this small example, compressed JSON is 2--3$\times$ the size of Avro. JSON attribute names are encoded in every record as key-value pairs to attribute values, meaning that the length of an attribute name can have a significant impact on file size. Attribute name storage is unnecessary, and is encoded by position in Avro records. 

We also exported \pname{} records in Parquet, which is a binary-encoded columnar storage format---it stores records together by column rather than row. This has several advantages for compression, reading, and data parsing that make it popular for big data applications.  Unfortunately, Parquet is not an ideal format for streaming data; however, it may be a viable serialization option once data has been pushed to cloud storage since it is almost as compact as Avro, and is optimized for query-based analytics. We plan to explore the efficacy of this data conversion going forward.

\section{Live Deployments}
\label{s:livedeploy}

\inlinedsectionit{Enterprise Portal}
We used the \pname{} stack to monitor a project management portal for a large organization with a thousand daily users and over 100 managed projects. The website is built using a set of docker containers and is composed of a front-end, back-end rest API, cache, and database. All incoming web traffic is received from an external gateway.  


The collector was deployed on the production server and \pname{} records were exported to an external S3-compliant object store server in 5-minute intervals. 
The front-end web server received over 50K web requests, and generated several million read and write operations while serving up pages.

A quick analysis of \pname{} traces provided a list of the different binaries that were executed on these production containers (see Table \ref{t:ldOpTable}). Of particular interest is the execution of a shell and several command line executables including \code{grep}, \code{jq}, \code{sed} and \code{tr}. The execution of these command line programs corresponded to a user entering one of the containers from the local host. The traces also reveal the arguments of each command and how they were used in the system.  


\begin{table}[t]
\centering
\scriptsize
\caption{Summary of a production environment telemetry.}
\begin{tabular}{@{}p{2cm}@{\hskip0.15cm}p{6.4cm}}
\toprule
{\footnotesize\textbf{Feature}} & {\footnotesize\textbf{Findings}} \\ 
\midrule
{\footnotesize{Executed binaries (names only)}} & redis-server, 
apollo-engine-binary-linux, 
node, npm, nginx, java, logspout, sh, sed, tr,  
python3.7, grep, jq, \ldots \\ 
\cmidrule{2-2}
{\footnotesize{PE operations (\#)}} & \tcode{clone}: 9421, \tcode{exec}: 4585, \tcode{exit}: 7277, \tcode{setuid}: 1041 \\
\cmidrule{2-2}
{\footnotesize{FE operations (\#)}}  & \tcode{mkdir}: 5837,
\tcode{symlink}: 4070, \tcode{unlink}: 1740,
\tcode{rename}: 906 \\
\cmidrule{2-2}
{\footnotesize{FF opflags (\#)}} & \tcode{open}: 3517723, \tcode{read}: 3676020, \tcode{write}: 127930, \tcode{close}: 3629040 \\
\cmidrule{2-2}
{\footnotesize{NF opflags (\#)}} & 
\tcode{accept}: 169589, \tcode{connect}: 897620, \tcode{shutdown}: 46043, \tcode{recv}: 1360250,
\tcode{send}: 1201979, \tcode{close}: 1267324 \\ 
\bottomrule
\end{tabular}
\label{t:ldOpTable}
\end{table}

\inlinedsectionit{Cloud DevSecOps}
In another setting, we deployed \pname{} on a twelve-node Kubernetes cluster to monitor a web-based enterprise management service hosted on a cloud environment. Our data processing pipeline (\S\ref{s:language}) was configured with a policy comprising 30 MITRE ATT\&CK TTPs, including interactions with sensitive files, process executions, and suspicious network activity. As part of routine security tests and hardening, an external penetration testing team automated scripts to probe and modify the service's containers, and setup command-and-control (C\&C) channels between the containers. We used the collected telemetry to assess both the security posture of the container environment and the development practices that went into building and configuring the containers. 

Our analysis uncovered critical security issues with this deployment. For example, our data showed that an ephemeral container was used to export data from one database to another at intervals. Each run, this container re-installed all private keys into key stores, and used clear-text connection strings to authenticate to the databases. Moreover, although the main web server container was executed under an unprivileged user, that user still had write access to modify the application's web files, and compile source files using a \code{javac} binary left in the unsanitized container image. We were also able to detect the C\&C communication patterns between the containers. Our system telemetry stack enabled us to effectively automate the extraction of security-relevant TTPs and incorporate runtime monitoring and attack vector discovery as a step of the development and security operations (DevSecOps) workflow.

\section{Conclusion}
\label{s:conclusion}
\pname{} provides deep visibility into computer workloads and enables scalable system analysis. It lifts raw system call information into a higher-level representation that captures process and container behaviors and interactions with other resources on a system. Evaluations on a wide variety of container-based workloads show that \pname{} reduces the telemetry footprint by several orders of magnitude over existing state-of-the-art solutions. Its open format and architecture makes it suitable for the creation of a common platform for system telemetry and analytics that will afford practitioners time to focus on high-value tasks rather than infrastructure building.

\bibliographystyle{IEEEtran}
\bibliography{bibliography}

\end{document}